\documentclass[10pt, a4paper, twocolumn]{article}
\usepackage{graphicx}%
\usepackage{multirow}%
\usepackage{amsmath,amssymb,amsfonts}%
\usepackage{mathrsfs}%
\usepackage[title]{appendix}%
\usepackage{xcolor}%
\usepackage{textcomp}%
\usepackage{manyfoot}%
\usepackage{booktabs}%
\usepackage{algorithm}%
\usepackage{algorithmicx}%
\usepackage{algpseudocode}%
\usepackage{listings}%

\usepackage[
    sorting=none,
    style=nature,
    url=true,
    isbn=false,]{biblatex}
\AtEveryBibitem{
  \clearlist{language}
  \clearfield{language}
}
\graphicspath{{figures/}{./}}

\begin{document}
\twocolumn[
\begin{@twocolumnfalse}
\begin{center}
   {\LARGE \bfseries Threat Evasion and Information Propagation in Cognitive Flocks \par}    \vspace{1em}
    \normalsize
    Priyanka Iyer$^{1}$, Cecilia Soroco$^{1,2}$, and Gerhard Gompper$^{1,\dag}$ \par
    \vspace{0.5em}
    \small
    $^1$Theoretical Physics of Living Matter, Institute for Advanced Simulation, Forschungszentrum Jülich, D-52425 Jülich, Germany \\
    $^2$ University of British Columbia, Vancouver, BC V6T 1Z4, Canada \\
    \vspace{0.5em}
    $^\dag$Corresponding author. Email: g.gompper@fz-juelich.de\par
    \vspace{1.5em}
\end{center}
\begin{abstract}
Information sharing, threat perception, and collective evasion highlight the advantages of swarm intelligence. Here, we study the behavior of flocks of cognitive agents, based on numerical
simulations of the augmented inertial spin model,  
with an explicit predator. Each agent on the
predator-facing side of the flock perceives the threat, triggering collective turns through
local interactions. We show that the augmented inertial spin model sustains nearly linear information propagation in the under-damped regime, even with realistic flocking interactions such as local collision avoidance and swarm cohesion in addition to alignment.  The turning distance of the flock to a stationary predator scales with the ratio $\Delta_p/(v_0\sqrt{\eta})$, identifying an optimal balance between response strength $\Delta_p$, flock speed $v_0$, and damping $\eta$ of information propagation. Fragmentation arises when different parts of the flock turn in different directions, however, fast information propagation can synchronize these responses and maintain flock cohesion while under pursuit. For two predators, a combination of stability analysis and simulations reveals a stark contrast between flock and single-particle reactions, with scattering angle shaped by flock geometry, predator position, and the heterogeneous responses of the detecting front. Together, these results show how information propagation, flock geometry, and dynamics together shapes collective escape, providing general principles relevant to both biological flocks and robotic swarms.
\end{abstract}
\vspace{1cm}
\end{@twocolumnfalse}
]

\section{Introduction}
In nature, collectives of relatively simple organisms, such as bees, ants, fish, or birds, are capable of collectively performing complex and coordinated tasks. 
Bees waggle to indicate direction and distance to flowers \cite{seeley_honey_1994}, ants release pheromones to determine optimal routes to food resources  \cite{bert_ants_1990}, fish and birds engage in fast collective turns to avoid threats such as predators \cite{papadopoulou_mechanistic_2026}, and African buffalo are found to `vote' to achieve a consensus on which direction to move \cite{prins_selecting_1996}. 
Fireflies have been found to synchronize their flashes through local interactions that propagate across the swarm in a relay-like manner \cite{sarfati_self-organization_2021}. 
Even at the microscale, fast information transfer guides a variety of biological processes, which cannot be achieved via simple processes such as the diffusion of molecules. Wound healing relies on the fast transmission of wound-induced cell-to-cell H$_2$O$_2$ signals, and the nervous system functions on the rapid transmission of electrical signals across neurons. Coordinated cell division, known as “mitotic waves” in eggs of insects and amphibians, and leaf closing in \emph{Mimosa pudica} also rely on fast cell-to-cell information transfer \cite{awan_characterizing_2018}. Such surprisingly complex behaviours arise from simple local interactions within members of these groups. A central question in several of these systems is the concept of information propagation. Here, `information' is a rather unspecific notion, which can refer to something as simple as a directional change of the group, without real knowledge of the nature of the threat itself. Fast, robust information transfer is essential for group cohesion, while at the same time, the group's response strategy should be resilient to minor, non-threatening heading changes caused by individual movement noise \cite{olsen_information_2026}. 


Collective escape is an important biological function seen in nature. Several experimental studies have characterised collective escape among different animal groups. At specific predator attack angles, schooling sardines exhibit a “fountain” escape, in which they split into two subgroups that arch around the predator and rejoin behind it \cite{bartashevich_collective_2024}. Controlled attacks on pigeon flocks by a robotic falcon demonstrate that escape frequency increases as the predator approaches, even though individuals do not prioritise escape any differently \cite{papadopoulou_self-organization_2022}. In predator–prey pairs such as lion–zebra or cheetah–impala, where predators are marginally faster than their prey, the prey can compensate for their lower speed with greater maneuverability \cite{wilson_biomechanics_2018}. Interestingly, this strategy mirrors findings from microscale studies, where simple run-and-tumble escape behaviours can improve survival by performing a sudden tumble when very close to the predator \cite{goh_adaptive_2026}. The nature of prey escape also depends on the type of hunting style and the perceived risk of the predator \cite{palmer_reactive_2021}. 

While species-specific approaches \cite{sayin_behavioral_2025,gelblum_ant_2015,gomez-nava_intermittent_2022} helped to uncovered new inter-agent sensing mechanisms and agent hierarchies, their diversity also implies that species-specific solutions may be unable to identify the general principles that govern collective escape. This motivates the use of minimal agent-based models to study predator-prey systems and general escape strategies \cite{demsar_simulating_2015,mohapatra_behavioural_2025,negi_binary_2025}.  In this work, we focus on the fast collective turns observed in fish schools and bird flocks, which typically arise from sensing and responding to local changes in heading direction. Birds generally interact with only 6–8 neighbours \cite{bialek_statistical_2012}, highlighting that collective turns emerge through the propagation of information across the group. Experiments with both birds and fish indicate that this information spreads rapidly in a wave-like manner \cite{procaccini_propagating_2011,attanasi_information_2014,gerlotto_waves_2006,herbert-read_initiation_2015}, in contrast to the diffusive scaling predicted by the Vicsek model, which has been widely used to study flocking and collective behaviour \cite{ginelli_physics_2016,negi_collective_2024,barberis_large-scale_2016}. Several minimal models have been proposed to explain how such fast, linear responses are achievable. These include the copying of agitation waves, in which a maneuver is sequentially repeated throughout the group \cite{hemelrijk_what_2015}, and a swarm model in which individuals switch between stationary and spinning states based on the body orientation of their neighbours \cite{zheng_body_2024}. Incorporating inertia into the orientation dynamics of agents likewise permits linear information propagation, analogous to the “spin waves” observed in magnetic materials --- a mechanism captured by the inertial spin model (ISM) for bird flocks \cite{attanasi_information_2014,cavagna_physics_2018}. The Vicsek model is in fact the overdamped limit of the ISM.
Also, a `time delay' in the interaction has been considered, i.e. retarded reactions by the agents that can be attributed to natural delays between sensing and reaction \cite{holubec_finite-size_2021}. 

Here, we employ the augmented ISM~\cite{iyer_dynamics_2026}, which includes local collision avoidance and swarm cohesion in addition to alignment, to study how realistic behaviours, such as vision-based threat avoidance, govern prey responses. This formulation provides the flexibility to recover Vicsek-like dynamics by taking the overdamped limit. In particular, we systematically investigate how factors such as speed, damping, and perceived threat from a predator shape the collective escape of the flock.

\section{Results}

\subsection{Model and Simulation Approach}

\textbf{Flock model:}
\noindent Each agent in the flock is modelled as an ``intelligent" active Brownian particle (iABP) \cite{iyer_emergent_2026} of mass $m$, which experiences a propulsion force $f_p$ acting along an orientation vector $\mathbf{e}_i$ and a frictional interaction with the embedding medium (with translational friction coefficient $\gamma$), so that the particle motion is determined by
\begin{equation}
\label{eq:motion}
	m \ddot{\mathbf{r}}_i =f_p\mathbf{e}_i-\gamma \dot{\mathbf{r}}_i.
\end{equation}
We operate in the overdamped limit, i.e large $m/\gamma$, so that $\dot{\mathbf{r}}_i\simeq f_p\mathbf{e}_i/\gamma$, where $f_p/\gamma=v_0$ is the particle speed. 
The dynamics of the orientation vector $\mathbf{e}_i$ is governed in the augmented ISM \cite{iyer_dynamics_2026,cavagna_flocking_2015} given by
\begin{equation}
\label{eq:steering}
  \begin{aligned}
      & \dot{\textbf{e}}_i =  \textbf{s}_i \times  \textbf{e}_i \\
      & \chi \dot{\textbf{s}}_i = \textbf{e}_i \times \left [ -\eta \textbf{s}_i + \boldsymbol{\zeta}_i + \textbf{M}_i \right ]
  \end{aligned}
\end{equation}
The spin $\textbf{s}_i$ associated with each particle $i$ acts to rotate the orientation vector $\textbf{e}_i$. The spin represents a generalized
momentum, $\chi$ a generalized rotational inertia, and $\eta$ the damping coefficient. The noise term $\boldsymbol{\zeta}_i$ is a Gaussian random process with zero mean and variance $\langle \boldsymbol{\zeta}_i (t) \boldsymbol{\zeta}_j (t') \rangle = (2d) \eta T \delta_{ij} \delta(t-t')$, where $T$ is a temperature that characterizes the noise in the rotation dynamics and $d$ is the dimensionality. 

All interactions with the neighbouring particles are incorporated in the generalized force $\textbf{M}_{i}$ \cite{iyer_dynamics_2026}, where
\begin{equation}
\label{eq:generalized_force}
\begin{aligned}
    \textbf{M}_i  & =  \textbf{M}_{align} + \textbf{M}_{avoid} + \textbf{M}_{follow} \\
    & +\textbf{M}_{threat}
\end{aligned}
\end{equation}


Here, $\textbf{M}_i$ points in the direction of the agents intended motion. The first three terms are the interactions within the flock.
$\textbf{M}_{align}$ incorporates the alignment interactions with strength $K$, $\textbf{M}_{avoid}$ governs local collision avoidance between neighbours with strength $\Omega_a$, and $\textbf{M}_{follow}$ is the cohesive term that drives the agents within the flock to regions of high density with strength $\Omega_f$. In addition, the cohesive interactions are assumed to depend on the local agent density [with a maximum strength of $\Omega_{f0}$], with a high $\Omega_f$ for agents with no neighbours and small $\Omega_f$ for agents surrounded by agents. The three interactions have ranges $R_a<R_k<R_f$, where $R_a$ is the avoidance range, $R_k=2R_a$ the alignment range and $R_f=3R_a$ the cohesion range. 
For each interaction term in equation~\eqref{eq:generalized_force}, there is an associated vision angle, which defines the spatial region within which the term acts. For simplicity, all vision angles are set to $\psi_\alpha = \pi$, i.e. panoramic view. For details see \emph{Methods}. These three interactions determine a minimal flocking model of an isolated flock. 
We set $\chi=1$ and $R_a=1$ in all simulations. Thus, the avoidance range $R_a$ sets the length scale in our simulations. The speed of linear information propagation in the ISM is given by \cite{cavagna_bird_2014}
\begin{equation}
    c_s = R_a \sqrt{K/\chi}, 
    \label{eq:speed}
\end{equation}
and is used to define the timescale $t_0=\sqrt{\chi/K}$.  \\

\textbf{Predator Perception and Avoidance:}
\noindent To study information propagation within the flock, we require that the agents be capable of detecting the predator. Most previous studies of information propagation in flocks, involving the original ISM \cite{cavagna_flocking_2015},  have focused on largely artificial responses, i.e., where a randomly selected agent is perturbed externally \cite{cavagna_flocking_2015,loffredo_collective_2023}. Here, however, we want to study a realistic threat response to an explicit predator --- stationary or mobile --- i.e. threat detection as an inherent property of each agent. Thus, we incorporate threat detection and avoidance as an additional interaction $\textbf{M}_{threat}$ of a similar functional form as the collision avoidance in Eq.~\eqref{eq:generalized_force} -- see {\em Methods} -- however with an initial threat perception only by agents on the predator-facing side of the flock, as the visual perception of other agents is {\em blocked} by neighbors.

\begin{figure}[t]
    \centering
    \includegraphics[width=0.95\linewidth]{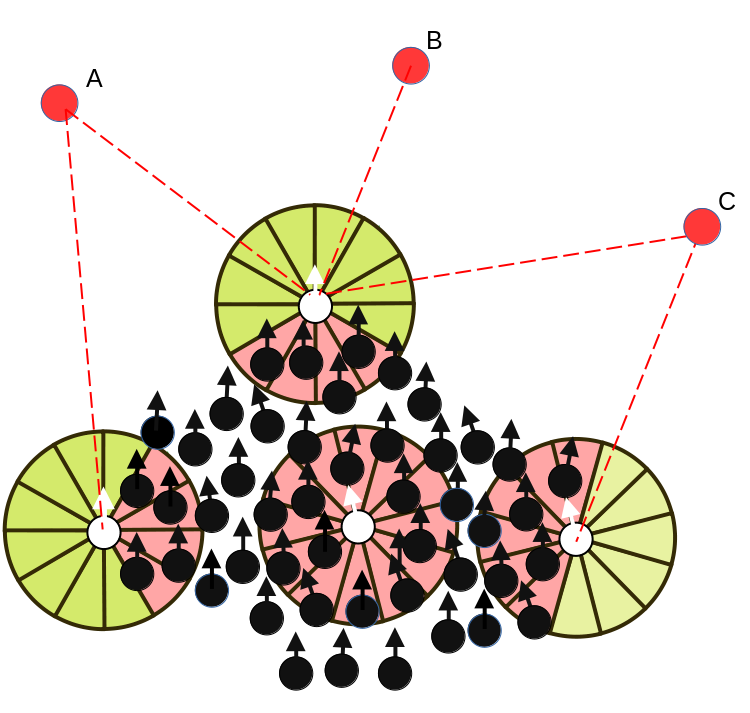}
    \caption{\textbf{Local vision occlusion}. Schematic of the local vision occlusion for selected particles (shown in white). Each vision cone (in two spatial dimensions) is divided into 12 segments, and if a segment contains another particle (shown in black), it is considered blocked (red shading). As a result, the threats (A, B, and C, red) are visible to only a few particles (connected via dashed-lines). The particle at the center of the flock has its entire field of view obstructed and cannot perceive any of the threats. The agents are modeled as point particles; the circles are shown for illustration only. }
    \label{fig:blocking}
\end{figure}

Thus, the \emph{primary} reaction of the flock to the perception of the predator is the initiation of a turning maneuver, with strength $\Omega_{p,l}$, by the agents with an unobstructed line of sight of the predator. The threat information should then propagate through the flock via social interactions (here alignment) \cite{jadhav_tracing_2026}. To mimic this visual obstruction, we construct a basic model of visual perception for systems in two spatial dimensions,  where each agent's field of view is divided into 12 equal angular segments, see Fig.~\ref{fig:blocking}. 
If a particular segment contains another agent within the alignment distance $R_k$, then the perception in this direction (segment) is considered blocked --- i.e. a predator in the same vision segment cannot be detected by the agent. With this basic implementation of realistic threat detection, each agent independently detects the predator based on its own perception, without any artificial or external selection of detecting agents. 

The \emph{secondary} reaction is assumed to occur when the predator approaches an agent at short distances $r < R'_{threat}=2 R_a$, with maneuverability $\Omega'_{p,s}>\Omega_{p,l}$. This interaction implies the triggering of a very strong avoidance steering when the threat is very close to the agent. In this case, no visual occlusion is implemented as the threat is already close to the agent, so that it is always detected. We define the relative threat response $\Delta_p = \Omega_{p,l}/K$, which quantifies the \emph{primary} threat response relative to the alignment strength $K$.  We operate in two spatial dimensions, which simplifies the equations of motion, see {\em Methods}. Unless specified otherwise, the simulations correspond to flocks with $N=400$ agents, $K=40$, and $\Omega_a/K=0.1$, which corresponds to the a state deep the polar-aligned phase \cite{iyer_dynamics_2026}. For a full list of simulation parameters, see Table~\ref{table:sim_para}.
 

\subsection{Characterizing Information Propagation}

An important first question to address is whether the presence of collision avoidance and 
swarm cohesion in the augmented ISM still allows for fast, undamped, linear information propagation, as the equations of motion differ significantly (in complexity) from the original ISM with only alignment interactions. Also, we will elucidate the effect of damping coefficient $\eta$ and the flock velocity $v_0$ on the information propagation within the flock.

To address these issues, we perform simulations of a simple scenario: a flock moving directly toward a stationary threat.
To check whether the bulk of the flock supports wave-like information propagation, we calculate and analyse rank curves, which measures the distance $x(t)$ versus the time $t$ taken for information 
about the threat (e.g.~\emph{turning/heading}) to propagate through the flock (see Fig.~\ref{fig1:rank_curves}).  For details on the calculation of rank curves, see Supplementary Information (SI), Sec.~S1.

\begin{figure*}
    \centering
    \includegraphics[width=0.99\linewidth]{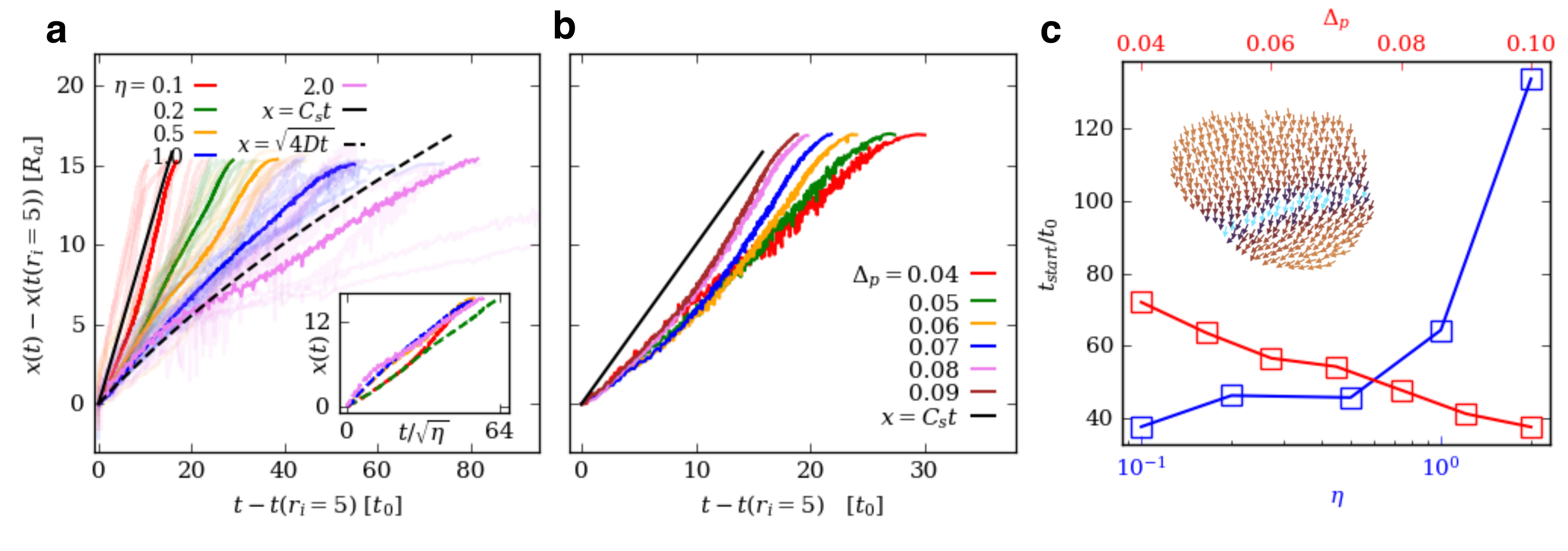}
    \caption{\textbf{Information propagation in the flock.} (a) Rank curves for varying damping coefficients $\eta$. For small $\eta$, the information propagates linearly (black solid line) whereas for large $\eta=2$, a diffusive propagation develops (black dashed line). Here the curve for the linear limit is shifted by the intercept at the start-time as $c_st-c_st_{start}(\eta=0.1)$, whereas the diffusive limit (black dashed line) is shifted by $\sqrt{4Dt}-\sqrt{4Dt_{start}(\eta=2)}$. The start time $t_{start}$is defined as the time taken for the initiators to turn by more than $\pi/4$ relative to their initial heading (here the first five agents).
    {\em Inset:} As $\eta$ increases, diffusive propagation develops resulting in a reduction in the slope as $1/\sqrt{\eta}$. 
    (b) Rank curves for fixed $\eta=0.1$, $v_0=0.2$, and varying predator response $\Delta_p$. 
    In (a) and (b), the curves are plotted for the ranks $r_i>5$, i.e. excluding the initiators of the turn.
    (c) The start-time $t_{start}$ of the turn for various $\Delta_p$ and $\eta$. 
    {\em Inset:} Snapshot of the propagating information wave (blue), which separates the leading front which has already turned (bottom) from the unperturbed, still unaware of the threat, (top) part of the flock. See Movie M1 for visualization of the information wave. }
    \label{fig1:rank_curves}
\end{figure*}

For a flock in the underdamped limit, information is expected to propagate with speed $c_s$, see Eq.~\eqref{eq:speed}.
Figure~\ref{fig1:rank_curves} shows the rank curves for various values of $\eta$ and $\Delta_p$. For the given flock parameters, the underdamped regime is approximately given by $\eta < 0.5$. Indeed, Fig.~\ref{fig1:rank_curves}(a) confirms that for small $\eta$, information propagates linearly with the expected speed $c_s$ (black solid line). This demonstrates explicitly that the flock can propagate information even when more realistic elements of cognitive self-steering --- such as flock cohesion, collision avoidance, and threat detection --- are incorporated in the model. Note that if the cohesion interactions are too strong in the bulk, this behaviour is not expected to hold, as the rotational symmetry of the ISM-Hamiltonian is broken. Additionally, for increasing perturbation strength, i.e., relative threat response $\Delta_p$, there is a marginal increase in speed until it saturates toward the prediction of Eq.~\eqref{eq:speed} (black solid line) in Fig.~\ref{fig1:rank_curves}(b) --- consistent with expectations, as the propagation speed is determined by the intrinsic properties of the flock (such as alignment strength, damping, and local interactions) and not by the strength of the initial response.

The start time $t_{start}$, defined as the time taken for the initiators to turn by more than $\pi/4$ relative to their initial heading (here the first five agents), increases with $\eta$ (see Fig.~\ref{fig1:rank_curves}(c)), and decreases with increasing $\Delta_p$. The decrease in $t_{start}$ with increasing perturbation strength $\Delta_p$ occurs because the nearest neighbours of the alert agents only react when the heading of the alert agent changes beyond a certain threshold, which happens more quickly for stronger perturbations. Increasing the damping coefficient $\eta$ dampens out the initial perturbation, leading to a larger start time $t_{start}$. In fact, in the overdamped limit, the effective interaction strength decreases as $\Omega_{p,l}/\eta$. 

\begin{figure*}
    \centering
    \includegraphics[width=0.99\linewidth]{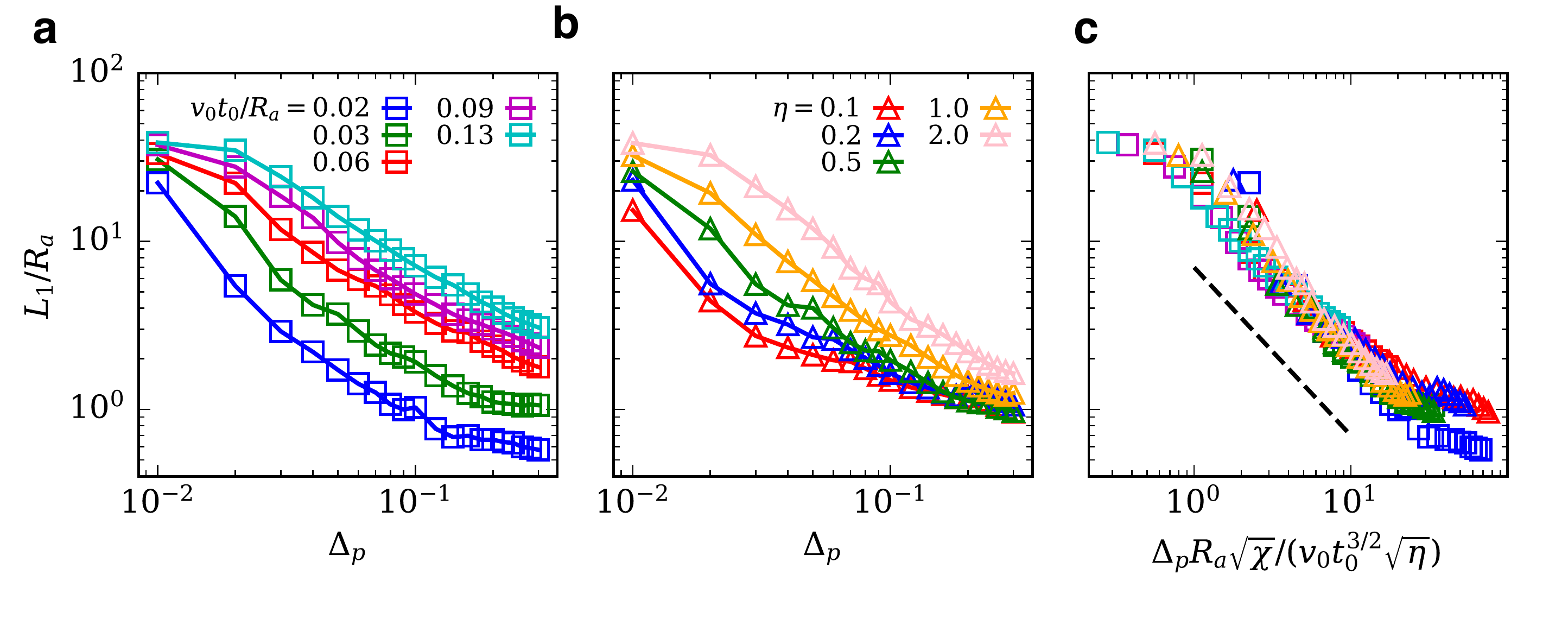}
    \caption{\textbf{Turning distance of the flock after perception of a predator.} The turning distance $L_{1}/R_a$ as a function of the predator-response strength $\Delta_p$, for 
    (a) different flock velocities $v_0$ ($\eta=0.5$) and 
    (b) different $\eta$ (Here $v_0/c_s=0.03$). 
    (c) A universal scaling curve is obtained when data for $L_{1}$ from (a) and (b) is plotted against the scaling variable $\Delta_p R_a \sqrt{\chi}/(v_0 t_0^{3/2} \sqrt{\eta})$. 
    The dashed line indicates the decay $L_{1}/R_a \sim (v_0 t_0^{3/2} \sqrt{\eta}) / (\Delta_p R_a \sqrt{\chi})$.}
    \label{fig2:closest_app}
\end{figure*}

In the deeply overdamped regime, for $\eta=2$, we find that the rank curve approaches $x(t)=\sqrt{4Dt}$, with $D=KR_a^2/\eta$, i.e. the diffusive propagation expected in the Vicsek limit. At intermediate values of $\eta$, we observe that the propagation remains linear or nearly linear (for $\eta=1$), but the slope of the rank curve decreases as $\sqrt{D}\sim 1/\sqrt{\eta}$ (see Fig.~\ref{fig1:rank_curves}(a) inset). This indicates that a diffusive response kicks in as $\eta$ increases. These results show that our model captures both diffusive and ballistic (linear) transport regimes. This implies that the inclusion of vision-based avoidance and cohesion has no adverse effect on the flock's overall ability to respond rapidly: the augmented ISM reproduces both types of propagation. The inset in Fig.~\ref{fig1:rank_curves}(c) shows the propagation of the turn in the flock, with the wave front separating the leading front that has already turned from the rear part that is still unaware of the threat.

\subsection{Turning Distance after Predator Perception}

We examine the turning distance $L_{1}$, defined as the distance travelled in the direction of the threat before the turn is completed.  As discussed above, increasing threat-response strength $\Delta_p$ leads to a smaller turning distance $L_{1}$. With increasing $v_0$, a 
larger $\Delta_p$ is required to maintain a short turning distance, see Fig.~\ref{fig2:closest_app}(a).  This bears some similarities to pedestrian experiments and simulations, where higher pedestrian velocity leads to closer encounters \cite{negi_controlling_2024,echeverria-huarte_estimating_2021}. A similar requirement of a larger $\Delta_p$ to maintain a short distance between a flock and the predator is also seen also for increasing $\eta$ [Fig~\ref{fig2:closest_app}(a)]. All data collapse onto a single master curve with
\begin{equation}
    L_{1}/R_a = \Gamma \left(\Delta_p R_a \sqrt{\chi} / (v_0 t_0^{3/2} \sqrt{\eta})\right ),
\end{equation}
with scaling function $\Gamma$, where $\Gamma(x)\sim 1/x$ over a wide range of the scaling variable, see Fig~\ref{fig2:closest_app}(c). Moreover, we find that for scaling variable $x>10$, the turning distance $L_1$ saturates. 
This indicates the existence of a minimum `activation' threshold, 
$\Delta_p R_a \sqrt{\chi} / (v_0 t_0^{3/2} \sqrt{\eta}) \simeq 10$, beyond which the turning distance of flock does not change significantly -- suggesting an optimal value of $\Delta_p$ that the flock should maintain to ensure the most effective response. 
This can also be seen in the rank curves, Fig.~\ref{fig1:rank_curves}(b), where for $\Delta_p \gtrsim 0.08$ (corresponding to $\Delta_p/(v_0\sqrt{\eta})\gtrsim 13$), the rank curve reaches the linear response predicted from a lattice model, i.e. where the agents are not moving and their interaction network is fixed \cite{cavagna_physics_2018}.

\subsection{Flock Splitting}

\begin{figure*}
    \centering
    \includegraphics[width=0.99\linewidth]{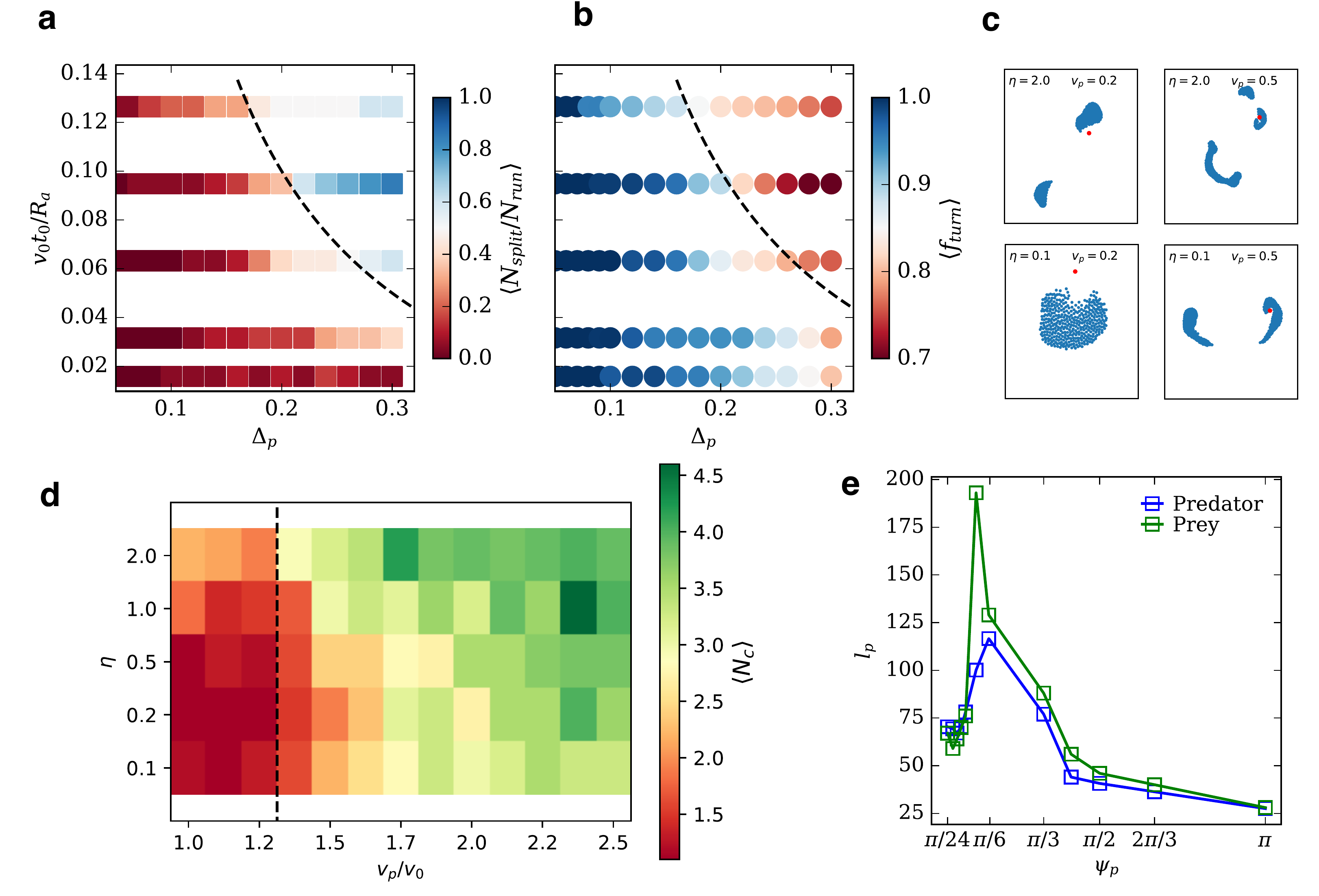}
    \caption{\textbf{Flock fracturing under predator attack.} 
    (a) Probability of flock splitting (measured from the fraction of simulations $N_{split}/N_{run}$ where the flock splits) as a function of scaled flock velocity $v_0 t_0/R_a = v_0/c_s$ and predator response strength $\Delta_p$. 
    (b) The fraction of agents $f_{turn}$ within each flock that initiate the turn in the same relative direction, i.e. left or right from the initial heading vector, as a function of $v_0$ and $\Delta_p$. The data is averaged over all simulation runs ($N_{run}=20$). Lines indicating phase boundaries are guides to the eye. 
    (c) Snapshots of the flock at time $t=2 t_L$ ($t_L = \sqrt{N}R_a / v_0$ is characteristic time required to traverse the flock size) for various predator speeds $v_p$ and damping coefficients $\eta$. 
    (d) Average number of clusters $\langle
    N_c\rangle$ at time $t = 2.5t_L$ after the onset of the attack, where $t_L = \sqrt{N}R_a / v_0$ denotes the characteristic time required to traverse the flock size. The results are averaged over 10 runs. The dashed line denotes the approximate threshold above which the predator will catch up to the flock, even if the flock performs a near-immediate turn at the start of the run. At the beginning of each run, the predator and the flock are oriented head-on toward each other. Here $v_0/c_s=0.03$.
    (e) Persistence length $l_p$ of the prey and flock particles as a function of the predator vision angle $\psi_p$.}
    \label{fig:split}
\end{figure*}

Depending on the flock parameters, the flock may split, or remain cohesive, upon approaching and turning away from the predator. To quantify this behaviour, we construct a phase diagram of flock splitting, see Fig.~\ref{fig:split}(a), which shows the fraction of simulations in which the flock splits (at the end of the run), as a function of flock velocity $v_0$ and threat response strength $\Delta_p$. For large $\Delta_p$ and/or $v_0$, the flock exhibits a higher tendency to split. Figure~\ref{fig:split}(b) displays the fraction of agents that initiates the turn in the same direction. This measure, computed when agents have turned by $\pi/2$ away from the predator (but the flock is still cohesive), represents the larger fraction of agents turning in the same direction (ranging from 0.5 to 1.0). It quantifies the degree of consensus the flock achieves midway through the collective turn, which depends on several factors, such as response speed, cohesion, and alignment dynamics. Note that even if the flock doesn't achieve consensus midway, it may still stay together at the end of the turn, due to alignment and cohesive interactions in the later stages of the turn.

There is a clear anti-correlation between the level of consensus achieved during the turn and the tendency of the flock to split, i.e., the lower the consensus, the higher the likelihood of splitting. At the onset of the turn, initiators are approximately evenly distributed to the left and right of the flock's heading vector from the centre of mass, so that roughly half attempt to turn to the left and half to the right. However, whether the flock actually splits depends on the perturbation strength, flock velocity, and the speed of information propagation within the flock. A stronger initial response --- i.e., higher perturbation strength --- increases the chance of splitting, as the initiators have an increasingly lower tendency to align with the overall flock direction, dividing the flock into two turning groups. At higher flock speeds, there is a competition between the motion of the agents and the speed at which information about the turn propagates \cite{loffredo_collective_2023}. 
As $v_0$ approaches the information propagation speed $c_s$, flock splitting obviously becomes unavoidable, consistent with the results in Fig.~\ref{fig:split}(a), as the flock splits, when the left- and right-turning groups diverge before they can influence each other through alignment and cohesion. 
We also expect the probability for flock splitting to increase with increasing (linear) flock size $L=\sqrt{N}R_a$. A critical flock size is determined by the distance $L^*$ a signal needs for travelling from one side of the flock to the other side during the turning time $t_1=L_1/v_0$, which is $L^*=t_1 c_s = L_1 c_s/v_0$. Much larger flocks may even fragment into several groups, as they turn away from the threat independently.

\subsection{Predator Approaching}

We now consider the scenario in which the predator approaches the flock with a nonzero speed $v_p > 0$, and thereby chases the flock through self-steering, for various damping coefficients $\eta$ and fixed threat response $\Delta_{p}$. See \emph{Methods} for the equation of motion of the predator. 
The predator is initiated at a distance $L=\sqrt{N}R_a$ from the centre of mass of the flock, i.e. a flock diameter away, with the predator and the flock oriented head-on toward each other. Figure~\ref{fig:split}(c) presents a few snapshots of the flock for various various speeds $v_p$ and damping coefficients $\eta$ [see SI Fig.~S1 for a full set of snapshots]. As $v_p$ increases, the predator penetrates deeper into the flock (measured at a fixed time after the onset of the turn), causing the group to split into several smaller sub-flocks, see Movie M2. Note that as the predator can approach the flock closely, the short-range (strong) interaction is triggered (i.e., $\Omega_{p,s}$), which causes the flock to split as each agent now prioritises escape over cohesion. Figure~\ref{fig:split}(d) shows a heat map of the number of clusters at time $t = 2.5 t_L$, where $t_L = L / v_0$ represents the characteristic time required (by the flock) to traverse its own size.
For marginally faster predators, with $v_p/v_0\gtrsim 1$, and small $\eta$ (fast turning), the flock has the lowest tendency to split, highlighting the importance of rapid information propagation in maintaining group cohesion. Even for a near-immediate turn, the earliest time the predator can reach the flock (for the given initialization, and under the assumption of a circular flock shape) is $\sqrt{N}R_a / [2(v_p - v_0)]$. Thus, for a simulation of length $t = n t_L$, this implies that for $v_p > v_0[1 + (2n)^{-1}]$, the predator will reach the flock within the defined simulation time. For $n = 2.5$, any $v_p/v_0 \gtrsim 1.2$ can cause the flock to split, as confirmed by the simulation data. Notably, even below this threshold, for large $\eta$, the flock still splits because the turning time is longer, allowing the predator to traverse part of the distance while the flock turns to avoid it.

As discussed above, for large $v_p$, the flock invariably breaks apart as the predator eventually catches up to the prey. This implies that fast turning alone is insufficient to evade predators with significantly higher velocities. Nonetheless, if we assume that predators can sustain such high speeds only in short time bursts, rapid turning may still serve as an effective short-term escape strategy. In general, however, when facing faster predators, prey must employ more sophisticated evasion tactics --- such as sudden direction changes or coordinated maneuvers to enhance survival \cite{goh_run-and-tumble_2025, wilson_biomechanics_2018}. However, sharp directional changes may be 
possible for individual prey, but not be feasible for a flock, which split into several smaller groups, as seen in fish schools under attack \cite{nottestad_herring_1999}. 

\begin{figure*}
    \centering
    \includegraphics[width=1.1\linewidth]{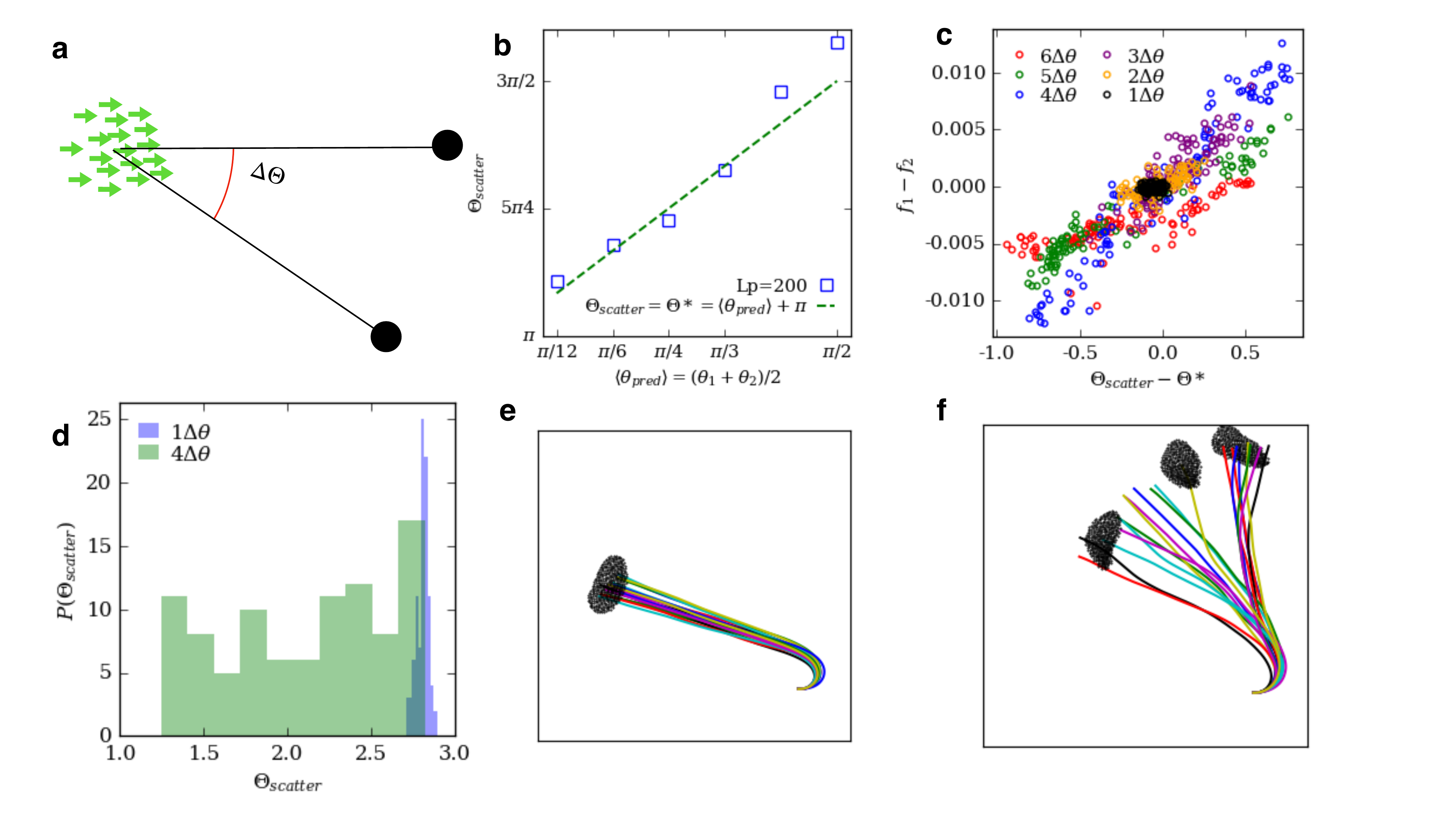}
    \caption{\textbf{Variability of flock escape from two predators.} (a) Schematic of the two-predator scattering simulation. (b) Exit angle $\Theta_{scatter}$ of the flock for two stationary predators at positions (radial coordinates) $(L_p,\theta_1)$, $(L_p,\theta_2)$, with $\theta_1=0$, i.e. on the x-axis. The dashed line indicates the exit angle equal to the fixed point solution for the average polarization direction $\Theta^*=(\theta_1+\theta_2)/2 + \pi$. (c) The difference $f_1-f_2$ between the fraction of boundary particles that only detect predator `1' ($f_1$) and predator `2' ($f_2$) as a function of the deviation $\Theta_{scatter}-\Theta^*$. In all simulations $R_{threat}=2L_p$ so that the threat is detected for the duration of the run.
    (d) Probability distribution of the scattering angle of flocks reacting to threats with $\Delta\Theta=1\Delta\theta$ and$\Delta\Theta=4\Delta\theta$, and (e,f) corresponding trajectories. Data are shown for 100 independent simulations. 
    In (e,f), some representative flock conformations at the end of the simulation are also shown. See SI, Fig.~S2,S3 for all trajectory data. }
    \label{fig:two_pred}
\end{figure*}

\subsection{Role of Predator Vision Angle}

We vary the vision angle of the predator in the range $\psi_p \in (\pi/24,\pi)$, capturing the full range from very focused to panoramic vision. 
The persistence lengths $l_p = v\tau$ of predator and flock are calculated from the characteristic decay of their velocity auto-correlation function $C(t)$, where $v = v_0$ for the flock and $v = v_p = 1.1 v_0$ for the predator, and the time $\tau$ at which the velocity auto-correlation has decayed to $C(\tau) = 1/e$. 
Figure~\ref{fig:split}(e) shows the results for the persistence lengths of flock and predator as the predator vision angle $\psi_p$ is varied. 
Firstly, both the flock and predator display essentially the same functional dependence. 
Since the flock is cohesive and tries to escape from the predator, it is clear that the predator movement guides the flock motion. In this sense, the predator is the driver and the flock behavior is reactive in nature, with long-term flock dynamics governed by the predator motion. 
Secondly, the persistence length of the predator peaks at $\psi \simeq \pi/6$, which is a narrow-to-intermediate vision angle. This implies that at this vision angle, the predator persistently follows some region of the flock, leading to a much higher (almost double) persistence length than that for narrow or wide vision angles. Here, very narrow vision angles imply that the predator can lose sight of a prey agent very quickly, as noise can bring other agents into the field of view and be selected as
a new target. In contrast, a wide vision angle implies that the predator perceives too many prey and is unable to persistently follow one subgroup. This qualitatively illustrates the \emph{confusion effect} that large flocks create to escape predators \cite{hogan_confusion_2017} -- which is an important advantage of swarm formation for predator escape.

\subsection{Flock Approaching two Predators}

To understand the effect of multiple predators, we consider a controlled scenario, where the flock reacts to two stationary predators. In this setup, the predators are placed at an angular separation of $\Delta \Theta = n\Delta\theta$, where $1 \leq n \leq 6$ and $\Delta\theta=\pi/6$. The flock is facing head on towards the first predator (see Fig.~\ref{fig:two_pred}(a)), such that for predator positions $(L_p,\theta_1)$ and $(L_p, \theta_2)$ in polar coordinates, the flock polarization at $t=0$ is along $(1,\theta_1)$. Here, the flock dynamics is overdamped, i.e., Vicsek-like. 

Simulation results, averaged over $100$ realizations for each $\Delta\Theta$, are shown in Fig.~\ref{fig:two_pred}(b). On average, the flock exits along the direction
\begin{equation}
\Theta^* = \frac{1}{2}(\theta_1+\theta_2) + \pi,
\label{eq:fixed_point}
\end{equation}
corresponding to motion along the perpendicular bisector of the line  between the two predators. However, a closer examination of the trajectories reveals a substantial spread in the exit angles, particularly for larger $\Delta\Theta$, see Fig.~\ref{fig:two_pred}(d) and Movie M3.

The origin of this variability can be understood in part from a stability analysis of the motion of a single agent (see SI Sec.~S2). For a particle displaced by a distance $\epsilon >0$ from the perpendicular bisector, the dynamics exhibit an instability that drives the particle even further away from the bisector, with the decreasing displacement rate with increasing distance $\epsilon$, see SI, Fig.~S5. Thus, the angle $\Theta^*$ of Eq.~\ref{eq:fixed_point} is actually an unstable fixed point. The flock displays a qualitatively similar behavior, effectively acting as an ensemble of particles initialized at different transverse positions. For the flock, an additional contribution is that boundary agents may only see one of the two predators due to vision blocking by other members of the flock, and of course flock cohesion. 

We therefore conjecture that the collective motion of the flock is governed by a {\em boundary majority mechanism}: the boundary region containing the largest number of predator-reacting particles dominates the response and determines the instantaneous scattering direction. This picture illustrates how the variance of $\Theta_{scatter}$ changes with $\Delta\Theta$. For {\em small} $\Delta\Theta$, most boundary particles detect both predators simultaneously. Consider a coordinate system with the bisector the $x$-axis, and the predator connection the $y$-axis. Particles located at $y>0$ tend to steer the flock toward positive $y$, while those at $y<0$ produce the opposite effect. Thus, despite $\Theta^*$ being an unstable fixed point for single-particle dynamics [see SI Fig.~S4(a,b)], the competing responses of the agents at the flock boundary nearly balance, driving its motion towards $\Theta^*$ with low variance, as shown in Fig.~\ref{fig:two_pred}(d,e).

In contrast, for {\em large} $\Delta\Theta$, the composition of the boundary becomes highly heterogeneous. Some particles detect only predator 1, others only predator 2, while a third group detects both predators. At the same time, the flock shape fluctuates dynamically, altering the relative abundance of these boundary populations. Consequently, small variations in flock geometry can bias the response away from one predator or the other. If the boundary region reacting primarily to predator 1 dominates, the flock exits with $\Theta_{scatter} < \Theta^*$; conversely, if the boundary reacting to predator 2 dominates, the flock scatters in the opposite direction. As before, particles that detect both predators contribute according to their position relative to the symmetry axis $y=0$. This implies a large variability in the exit angle, as seen in Fig.~\ref{fig:two_pred}(d,f) [compared to single-particle simulations, see SI Fig.~S4(a,c)].

To test this hypothesis, and derive quantitative predictions, we estimate the excess fraction 
$f_1-f_2$ of boundary particles responding predominantly to one of the predators, as a function of the scattering angle across all simulations (see SI, Sec.~S3). We find a strong correlation between $f_1-f_2$ and $\Theta_{\mathrm{scatter}}$, supporting the boundary-majority hypothesis, see Fig.~\ref{fig:two_pred}(c). The difference $f_1-f_2$ is mainly determined by the instantaneous flock shape and its orientation to the two predators (see SI Fig.~S6). Our results imply that a flock and individual (i.e., non-interacting) particles have very different scattering angles. The flock response is controlled by the competing influence of various predator-sensitive boundary regions, implying a highly nontrivial dependence on both predator perception and instantaneous asymmetry of the flock shape.

\section{Summary and Conclusions}

The formation of flocks and swarms is advantageous for the escape from predators.  Our simulation results indicate how the flock can optimize successful escape by choosing interaction parameters. We find that the approach distance is minimal for 
$\Delta_p R_a \sqrt{\chi} / (v_0t_0^{3/2} \sqrt{\eta}) \simeq 10$, with $\Delta_p=\Omega_{p,l}/K$. This implies that steering strength $\Omega_{p,l}$ in response to predator perception should be proportional to alignment strength $K$ and flock speed $v_0$. This result characterizes and summarizes several important properties of the flock. For smaller $\Omega_{p,l}$, the redirection and turning is too slow for an efficient escape. For larger $\Omega_{p,l}$, the information propagates not fast enough (propagation speed controlled by 
$\sqrt{K}$), and the flock splits and fragments. This also implies that flock splitting is inversely correlated with an early-turn consensus, as the alignment strength $K$ cannot  compete with the desire (quantified by $\Omega_{p,l}$) of agents at different parts of the leading edge of the flock to turn away from the predator direction as quickly as possible.

Flock splitting and fracture occurs when subgroups of the flock turn in opposite directions and fail to rejoin, which becomes more likely at higher speeds $v_0$ or threat-response strengths $\Delta_p$. This behaviour is strongly anti-correlated with the degree of early-turn consensus, which measures how uniformly agents choose a turning direction. If consensus is low --- due to a fast, strong initial perturbation or insufficient time for information to spread --- the flock splits before alignment forces can restore cohesion. Although we have not investigated the dependence of the flock size, it seems obvious that larger flocks split more easily, as the time of information propagation through the flock increases linearly with its size. For a motile predator,  rapid information propagation is important for maintaining group cohesion. However, even if the flock responds quickly, a (faster) predator is able to catch up and penetrate the flock, causing it to break apart into multiple groups and directions.

A variation of the predator vision angle $\psi_p$ reveals two key effects. First, the predator actively drives the flock, so that the escape dynamics are dictated by the predator motion. Second, a narrow-to-intermediate vision angle maximizes the predator’s persistence time, resulting in fewer directional changes. This aligns with the results of studies of predation of schooling fish, which suggest that predators often lock onto a single target during short bursts of motion \cite{major_predator-prey_1978,jolles_both_2022}. 
Finally, in a controlled scenario with two stationary predators, the flock behaves very differently from individual agents, with the scattering angle strongly affected by the flock shape, angular separation of the two predators, and the heterogeneity of the flock boundary, i.e. how many agents detect one or both threats. The heterogeneity in the perception angle at the boundary driving the escape strategy is an indication for
an optimal flock size for escape: large flocks are ponderous, small flocks are more agile. Additionally, if the flock fractures, it has the advantage of confusing a predator, like for Buridan's ass (a hypothetical donkey standing between two equally appealing bales of hay, unable to decide which to chose \cite{noauthor_buridans_2026}), as it cannot quickly decide which flock fragment to pursue further.

Our simulation results are obtained for surface-bound systems. Thus, microrobotic systems might be good candidates for experimental comparison and verification. From  the theoretical side, an extension to flocks in three spatial dimensions is obviously desirable.

\section{Methods}
Here we provide additional details of the model.

\subsection{Flock Dynamics}
\label{app:flock}

The dynamics of agents within the flock are incorporated in the generalized force $\textbf{M}_{i}$, where
\begin{equation}
\begin{aligned}
    \textbf{M}_i  & = \textbf{M}_{avoid}+ \textbf{M}_{align} + \textbf{M}_{follow} \\
    & +\textbf{M}_{threat}
\end{aligned}
\end{equation}

These terms have the following form:

\begin{eqnarray}
\textbf{M}_{avoid} &=& - \frac{\Omega_a}{N_a} 
       \sum_{j\in VC_a} \exp{(-r/r_a)} \hat{\textbf{r}}_{ij}  \\
\textbf{M}_{align} &=& \frac{K}{N_k} \sum_{j\in VC_k} \textbf{e}_j \\
\textbf{M}_{follow} &=& \frac{\Omega_f}{N_f}  
           \sum_{j\in VC_f} \exp{(-r/r_f)}\hat{\textbf{r}}_{ij} \\
\textbf{M}_{threat} &=& - \frac{\Omega_{p,l}}{N_{p,l}} \sum_{j\in VC_{p,l}} \hat{\textbf{r}}_{ij} \nonumber\\ 
& & \hspace{1true cm}  - \frac{\Omega_{p,s}}{N^i_{p,s}} \sum_{j\in VC_{p,s}} \hat{\textbf{r}}_{ij} 
\end{eqnarray}

\begin{table*}
\centering
\begin{tabular}{l c}
\hline
\textbf{Parameter} & \textbf{Value} \\
\hline
Behavioural inertia $\chi$ & 1 \\
Spin damping $\eta$ & 0.1-2.0 \\
Flock velocity $v_0$ & 0.05-0.5 \\
Alignment strength $K$ & 40 \\
Maximum cohesion maneuverability $\Omega_{f0}$ & $K$ \\
Local collision avoidance maneuverability $\Omega_{a}$ & 0.1$K$ \\
Threat avoidance maneuverability (long) $\Omega_{p,l}$ & 0.0$K$-0.3$K$ \\
Threat avoidance maneuverability (short) $\Omega_{p,s}$ & $0.5K$ \\
Collision avoidance range $R_a$ & 1 \\
Alignment range $R_k$ & 2$R_a$ \\
Cohesion $R_f$ & 6$R_a$ \\
Threat avoidance range (short) $R'_{threat}$ & 2$R_a$ \\
\hline
\end{tabular}
\caption{List of parameters and their values for the augmented inertial spin model.}
\label{table:sim_para}
\end{table*}

The terms describe the following actions:
\begin{itemize}
    \item \textbf{Local collision avoidance:} The term $\textbf{M}_{avoid}$ steers the agents to avoid collisions locally in the range $R_a$, and sets the length scale in our system. It has a decay range $r_a=R_a/3$
    \item \textbf{Alignment:} The term $\textbf{M}_{align}$ drives the agents to align with the local neighbours in the range $R_k=2R_a$. 
    \item \textbf{Cohesion:} The term $\textbf{M}_{follow}$ steers the agents towards regions of high density. The cohesion interactions act in the range $R_f = 6R_a$, with a decay range $r_f=R_f/3$. However, the cohesion  interactions are ‘switched off’ in bulk, implemented by a sigmoid function $\Omega_{f}(n_c)=\Omega_{f0} [1+\exp(-(n_c-n_0))]^{-1}$, where $n_c$ is the number of neighbours in its alignment zone and $n_0$ is a constant. 

    \item \textbf{Threat avoidance:} The term $\textbf{M}_{threat}$ incorporates short ranged ($R'_{threat}=2R_a$) and long ranged ($R_{threat}\gg R_a$) threat detection. The long range threat detection additionally has visual occlusion implemented (as described in the main text). Here there is no (exponential) decay term as there are typically fewer predator particles. 
\end{itemize}
Furthermore:
\begin{itemize}
    \item \textbf{Vision cones:} For each interaction there is a vision cone defined by 
    \begin{equation}
    VC_{\alpha} = \left\{\,j \mid \frac{\textbf{r}_{ij}}{|\textbf{r}_{ij}|}\cdot e_i \geq \cos\psi_{\alpha} \text{ and } |\textbf{r}_{ij}|<R_{\alpha} \,\right\}
    \end{equation}
    where the $\alpha$ indicates the index of the interaction. Note that each interaction \emph{may} have a different vision cone. Unless specified otherwise, $\psi_\alpha=\pi$ for all interactions. The long range threat detection has visual occlusion in addition to this vision cone. 

    \item \textbf{Normalization:} The normalization factors $N_f=\sum_{j\in VC_f} \exp{(-r/r_f)}$, 
    $N_a=\sum_{j\in VC_a} \exp{(-r/r_a)}$, $N_k=\sum_{j\in VC_k}$, $N_{t,l}=\sum_{j\in VC_{t,l}}$, and $N_{t,s}=\sum_{j\in VC_{t,s}}$ measure the effective number of agents in the vision cones, so that 
    the interactions are non-additive.
\end{itemize}

For surface-bound systems, two-dimensional polar coordinates can be employed, with 
$\textbf{e}_i = (\cos\theta,\sin\theta)$, so that the equations of motions reduce to 
\begin{eqnarray}
    \dot{\theta} &=& s_i \\
    \chi \dot{s}_i &=& \zeta_i -\eta s_i + \frac{K}{N_k} \sum_{j\in VC_k} \sin(\theta_j-\theta_i) \\
    &-& \frac{\Omega_a}{N_a} \sum_{j\in VC_a}  e^{-r/r_a} \sin(\phi_{ij}-\theta_i) \\ 
    &+&  \frac{\Omega_f}{N_f} \sum_{j\in VC_f}  e^{-r/r_f} \sin(\phi_{ij}-\theta_i) \\
    &-& \frac{\Omega_{p,l}}{N_{p,l}} \sum_{j\in VC_{p,l}} \sin(\phi_{ij}-\theta_i) \\
    &-& \frac{\Omega_{p,s}}{N_{p,s}} \sum_{j\in VC_{p,s}} \sin(\phi_{ij}-\theta_i) 
\end{eqnarray}
where $\phi_{ij}$ is the polar angle of the vector $\hat{\textbf{r}}_{ij}$. Here, the cohesion and collision/threat avoidance terms are all of the same basic form, i.e. $\propto \sin(\phi_{ij}-\theta_i)$, and mainly differ by their sign (turning away/toward) and their interaction range. 

\subsection{Threat Dynamics}
\label{app:threat}

Similar to the flock dynamics, the predator dynamics are given by 
\begin{equation}
\begin{aligned}
& m \ddot{\mathbf{r}}^p_i =\gamma v_p \mathbf{e}^p_{i}-\gamma \dot{\mathbf{r}}^p_i \\
      & \dot{\textbf{e}}^p_i =  \textbf{s}^p_i \times  \textbf{e}^p_i \\
      & \chi \dot{\textbf{s}}^p_i = \textbf{e}^p_i \times \left [ -\eta \textbf{s}^p_i + \boldsymbol{\zeta}_i + \textbf{M}^p_i \right ]\\
    & \textbf{M}^p_{i} = \frac{\Omega_{chase}}{N^i_c} \sum_{j \in VC_p} \hat{\textbf{r}}_{ij}\\
\end{aligned}
\end{equation}
where $\Omega_{chase}$ is the maneuverability with which the predator chases the flock, $v_p$ is the predator speed and $\psi_p=\pi$ is the vision angle of the predator. In polar coordinates,

\begin{eqnarray}
    \dot{\theta}^p &=& s^p_i \\
    \chi \dot{s}^p_i &=&  \zeta_i -\eta s^p_i \\ 
        &+& \frac{\Omega_{chase}}{N_p} \sum_{j\in VC_p} e^{-r/r_p} \sin(\phi_{ij}-\theta^p_i) 
\end{eqnarray}
where the sum $j\in VC_p$ runs over the flock particles in the vision cone of the predator, with vision angle $\psi_p$. 

The equations of motion for the flock and the threat are implemented within the LAMMPS framework \cite{thompson_lammps_2022,iyer_collective_2025}, and the time integration in our simulations is performed using the velocity-Verlet algorithm \cite{allen_computer_2017}.

\printbibliography

\section*{Data Availability}

The simulation code, data, and analysis scripts that support the results of this study are available at Zenodo \cite{iyer_collective_2025,iyer_data_2026}.

\section*{Author Contributions}
G.G.~designed the project. 
P.I.~and C.S.~performed the simulations and analyzed the data. 
P.I., C.S., and G.G.~discussed the results. 
P.I.~and G.G.~wrote the manuscript.

\section*{Competing Interests}

The authors declare no competing interests.

\end{document}